\documentclass[twocolumn,showpacs,preprintnumbers,amsmath,amssymb]{revtex4}
\usepackage{bm}

\begin{document}

\title{Hyperfine Level Splitting for Hydrogen-Like Ions due to Rotation-Spin Coupling}

\author{I. M. Pavlichenkov}
 \email{pavi@mbslab.kiae.ru}
 \affiliation{Russian
Research Center "Kurchatov Institute", Moscow, 123182, Russia.}

\begin{abstract}
The theoretical aspects of spin-rotation coupling are presented.
The approach is based on the general covariance principle. It is
shown that the gyrogravitational ratio of the bare
spin-$\frac{1}{2}$ and the spin-1 particles is equal unity. That
is why spin couples with rotation as an ordinary angular momentum.
This result is the rigorous substantiation of the cranking model.
To observe the phenomenon, the experiment with hydrogen-like ions
in a storage ring is suggested. It is found that the splitting of
the $1\,^2\!S_{1/2}, F=1/2$ hyperfine state of the $^{140}{\rm
Pr}^{58+}$ and $^{142}{\rm Pm}^{60+}$ ions circulating in the
storage ring ESR in Darmstadt along a helical trajectory is about
4.5 MHz. We argue that such splitting can be experimentally
determined by means of the ionic interferometry.
\end{abstract}

\pacs{04.20.+v, 32.10.Fn, 03.75.Dg}

\maketitle

The spin of elementary particles is a fundamental quantum property
which does not have the classical analogue. On the other hand, the
gravity (in the broad sense as a consequence of the curvature of
space-time which governs the motion of inertial objects) seems to
be a pure classical phenomenon. There has been considerable
theoretical interest in the coupling of such different physical
entities since the paper of Kobsarev and Okun \cite{Kob}.
Currently, three different spin-gravity interactions are
considered: gravitoelectric, gravitomagnetic and spin-rotation
couplings. The latter was predicted independently by Mashhoon
\cite{Mash} and Silverman \cite{Silv}. This coupling reveals the
rotational inertia of the spin. Up to now there is no direct
experimental evidence of the phenomenon, however it could be
verified by present high precision experiments. We refer to the
recent measurements of the electron capture decay (EC-decay) rates
in hydrogen-like (H-like) $^{140}{\rm Pr}$ and $^{142}{\rm Pm}$
ions performed on the storage ring (ESR) of GSI Darmstadt
\cite{Lit,Lit1}. The time-resolved spectroscopy employed there
allows to study the time evolution of \textit{a single}
radioactive ion. The objective of the present paper is to show
that such experimental technique offers an opportunity for
observation of spin-rotation coupling, because ionic spectra carry
unambiguous information about the space-time curvature at the
position of an ion.

An ion circulating in a storage ring may move along a helical line
winding around the main orbit. This is the result of betatron
oscillations of a stable ion orbit. Consider the inertial frame of
reference $K$ with the origin $O$ moving along the main orbit with
the velocity $V$ ($\beta=V/c=0.71$ for the ESR) relative to the
laboratory frame. Define $X^0=ct$ and the Cartesian coordinate
axes $X^1,X^2,X^3$, with the $X^3$ axis along the main orbit. In
the plane $X^1,X^2$ an ion moves in a closed trajectory which we
assume to be a circle of the radius $R$ centered on the main orbit
in the point $O$. We introduce frame $K'$ having the common origin
with $K$ and uniformly rotating with respect to $K$ with angular
velocity ${\bf\Omega}(0,0,\Omega)$. In this frame the electron
(nucleus) is characterized by position vector ${\bf r}_e$ (${\bf
r}_n$), and the ion center-of-mass is ${\bf R}_0(R_0,0,0)$.
\cite{KK'}.

It is well known that the Schr\"odinger equation for the H-like
ion in the rotating frame acquires in the Hamiltonian the
additional term $-\hbar(\bf\Omega\!\cdot\!\bf l)$, where $\bf l$
is the electron orbital angular momentum. This is the Coriolis
interaction which should split degenerated magnetic sub-states.
However in this approximation the lowest hyperfine state
$1S_{1/2}(F=1/2,M_F=\pm 1/2)$ of the $^{140}{\rm Pr}^{58+}$ and
$^{142}{\rm Pm}^{60+}$ ions remains degenerate since the total
angular momentum ${\bf F}={\bf S}+{\bf I}$ includes only the
electron ($\bf S$) and nuclear ($\bf I$) spins.

The spin-rotation coupling has been obtained in refs. \cite{Mash}
and \cite{Silv} by exploiting a simplified approach which known in
nuclear physics as cranking model \cite{deV}. Shortly afterward,
Silverman \cite{Silv1} extended this method to consider hydrogen
hyperfine splitting in a rotating reference frame. We will see
below that this method allows to find only the main term in the
rotation induced perturbations of the energy levels of a heavy
ion. In addition, for high-Z one-electron ions, account must be
taken of the relativistic effects. Therefore, we will consider
relativistic wave equations for electron, electromagnetic field
and nucleus in a non-inertial reference frame by using the general
covariance principle. The starting point of our calculations is
the metric tensor in the rotating frame $K'$,
\begin{equation}
g_{00}=-1  +({\bm\omega}\!\times{\bf r})^2,\
g_{0i}=g_{i0}=({\bm\omega}\!\times{\bf r})_i,\ g_{ij}=\delta_{ij},
\label{metr}
\end{equation}
where ${\bf r}$ is ${\bf r}_e(x^1,x^2,x^3)$ for electron or ${\bf
r}_n(y^1,y^2,y^3)$ for nucleus and ${\bm\omega}={\bf\Omega}/c$. We
will use Latin letters to denote three spatial indices 1,2,3 and
Greek letters for four indices.

{\it Electron.}---In the standard formalism \cite{Chap}, the
general covariant Dirac equation has the form
\begin{equation}
\left\{\hat\gamma^\mu(x)\left[\partial_\mu -\Gamma_\mu+
\frac{ie}{\hbar c}A_\mu\right]+ \frac{mc}{\hbar}\right\}\psi(x)=0,
 \label{dir}
\end{equation}
where $A_\mu\{-\varphi,\bf A\}$ is the vector potential in the
non-inertial frame which describes electromagnetic interaction of
the electron with the nucleus. The coordinate-dependent matrices
$\hat\gamma^\mu(x)$ ($x=x^\mu=ct',x^1,x^2,x^3)$ satisfy the
equation
\begin{equation}
\hat\gamma^\mu(x)\hat\gamma^\nu(x)+\hat\gamma^\nu(x)\hat\gamma^\mu(x)=2g^{\mu\nu}(x).
\label{matr}
\end{equation}
The vierbein vectors $b^\alpha_{\: \mu}(x)$ are used to obtain the
relationship between the curved matrices and the flat ones
$\hat{\gamma}_0=i\hat{\beta},
\hat{\gamma}_i=-i\hat{\beta}\hat{\alpha}_i$ ($\hat{\beta}$ and
$\hat{\alpha}_i$ are the Dirac matrices)
\begin{equation}
\hat\gamma^\mu(x)=g^{\mu\nu}b^\alpha_{\:
\nu}(x)\hat{\gamma}_\alpha.
\end{equation}
The four different vectors $b^\alpha_{\: \mu}$ (on the $\mu$
index) connect with the metric tensor
\begin{equation}
g_{\mu\nu}(x)=\eta_{\alpha\beta}b^\alpha_{\: \mu}(x)b^\beta_{\:
\nu}(x),
\end{equation}
where $\eta_{\alpha\beta}$ is the flat space-time metric tensor
$\eta_{\alpha\beta}=\rm{diag}(-+++)$. Vierbein indices $\alpha$
are lowered with $\eta_{\alpha\beta}$ while vector ones $\mu$ are
raised with $g^{\mu\nu}$. Finally, the Fock-Ivanenko matrices
(spinor affine connections) $\Gamma_\mu$ are expressed in term of
the vierbeins and matrix $\hat{\gamma}_\alpha$
\begin{equation}
\Gamma_\mu=-\frac{1}{4}\hat{\gamma}_\alpha\hat{\gamma}_\beta
b^\alpha_{\: \nu}g^{\nu\lambda}(\partial_\mu b^\beta_{\: \lambda}
-\Gamma^\sigma_{\mu\lambda}b^\beta_{\: \sigma}), \label{FI}
\end{equation}
where the affine connections (the Christoffel symbols),
\begin{equation}
\Gamma^\sigma_{\mu\lambda}=\frac{1}{2}g^{\sigma\nu}
\left(\partial_\lambda g_{\nu\mu}+ \partial_\mu g_{\nu\lambda}-
\partial_\nu g_{\mu\lambda}\right), \label{Chri}
\end{equation}
are determined according to \cite{Land}.

The inertial effects are incorporated into eq. (\ref{dir}) through
the matrices $\hat\gamma^\mu(x)$ and the connections $\Gamma_\mu$.
To find these values, we must calculate
$\Gamma^\sigma_{\mu\lambda}$ and $b^\alpha_{\: \mu}$. The former
obtained from the metric tensor (\ref{metr}) are
\begin{eqnarray}
\Gamma^0_{00}&=&\Gamma^0_{i0}=\Gamma^0_{ij}=0, \nonumber \\
 \Gamma^i_{00}&=&e_{ijk}\omega_jg_{k0},\
\Gamma^i_{j0}=-e_{ijk}\omega_k,\ \Gamma^i_{jk}=0. \label{chr}
\end{eqnarray}
The vierbein vectors are expressed through the tetrad vectors
$h_\alpha^{\; \mu}$ found in Ref. \cite{Chap} as follows
$b^\alpha_{\: \mu}=g_{\mu\nu}\eta^{\alpha\beta}h_\beta^{\; \nu}$.
They take the form
\begin{equation}
b^\alpha_{\: 0}=\delta^\alpha_{\: 0}+\eta^{\alpha i}
e_{ijk}\omega_jx^k,\quad b^\alpha_{\: i}=\delta^\alpha_{\: i}.
\label{vier}
\end{equation}
Substituting these results in (\ref{FI}) we get the spin
connections
\begin{equation}
\Gamma_0=\frac{1}{4}\hat{\gamma}_i\hat{\gamma}_j
e_{ijk}\omega_k,\quad \Gamma_i=0.
\end{equation}

For the problem involving energy levels we need the Hamiltonian
form of the Dirac equation
\begin{equation}
i\hbar\frac{\partial\psi}{\partial t'}=H\psi. \label{Dir1}
\end{equation}
By multiplying (\ref{dir}) on the left by
$\hat\gamma^0(x)/g^{00}(x)$, one finds after simple calculations
the electron Hamiltonian
\begin{equation}
H_e=c\hat{{\bf\alpha}}\cdot\!\left({\bf p}_e+ \frac{e}{c}{\bf
A}\right)+eA_0+\hat{\beta} mc^2 -\hbar{\bf\Omega}\cdot({\bf
l}_e+{\bf S}), \label{eeq}
\end{equation}
where ${\bf p}_e$ is the electron momentum operator and ${\bf
l}_e=({\bf r}_e\!\times{\bf p}_e)/\hbar$ is its angular momentum
operator with respect to $O$. This exact result was obtained by
Hehl and Ni \cite{Ni} for free Dirac particles by using the
instantaneous inertial rest frame instead of the pseudo Riemannian
space-time.

Rotation-spin coupling is incorporated in the last term of
(\ref{eeq}). It is seen that the spin couples with rotation in the
same way as the angular momentum $\textbf{l}$, that means that the
gyrogravitational ratio of a bare Dirac particle is equal unity.
Equation (\ref{eeq}) is in fact the relativistic generalization of
the well known law of the classical mechanics concerned with
non-inertial reference frame \cite{Landm}. The orbital angular
momentum is substituted by the total one ${\bf j}={\bf l}+{\bf
S}$. Since $\bf j$ is the generator of rotations, we can transform
the flat space-time Dirac equation from the frame $K$ to $K'$ by
means of the unitary operator
\begin{equation}
{\cal R}=exp[-i({\bf j}\cdot{\bf\Omega})t]. \label{trans}
\end{equation}
The resulting stationary Hamiltonian can be shown to take the form
(\ref{eeq}). Important shortcoming of this method used in nuclear
physics as the cranking model, is that the rotation (\ref{trans})
requires extended interval of time $t>1/\Omega$. But, according to
the hypothesis of locality, the rotation time must be negligibly
small compared to the acceleration time $1/\Omega$.

{\it Electromagnetic field.}---To find the vector potential
$A_\mu$ produced by the point charge $Ze$ and the magnetic moment
${\bf\mu}=\mu_Ng_I{\bf I}$ ($\mu_N$ and $g_I$ are the nuclear
magneton and gyromagnetic ratio respectively), we have to consider
the Maxwell equations in the rotating frame $K'$. The covariant
form of these equations and the Lorentz gauge condition are
\cite{Land}
\begin{eqnarray}
g^{\lambda\sigma}\nabla_\lambda\nabla_\sigma A_\mu-R^\nu_\mu
A_\nu=-\frac{4\pi}{c}J_\mu, \quad  \nabla_\mu A^\mu=0,
\label{maxw}
\end{eqnarray}
where $R^\nu_\mu$ is the Ricci curvature tensor vanishing for the
metric (\ref{metr}) and $J_\mu=g_{\mu\alpha}J^\alpha$,
$J^\alpha\{cZe\delta({\bf r}-{\bf r}_n),{\bf i}\}$ is the current
vector. Using the covariant differentiation, and working to the
first order in $\Omega/c$, we find the following equations for the
vector potential
\begin{eqnarray}
\delta^{ij}\partial_i\partial_jA_0&=&4\pi\rho+
2e_{klm}\omega_k\partial_lA_m+
   \frac{4\pi}{c}e_{klm}\omega_ki_lx^m, \nonumber \\
\delta^{ij}\partial_i\partial_jA_k&=&
-\frac{4\pi}{c}i_k- 4\pi\rho e_{klm}\omega_lx^m,\nonumber \\
\delta^{ij}\partial_iA_j&=&-e_{klm}\omega_kx^l\partial_mA_0.
\end{eqnarray}
The solutions of these equations for a point nucleus are
\begin{equation}
A_0=-\frac{Ze}{\rho}- \frac{2({\bf\omega}\cdot\!{\bf\mu})}{\rho},
\quad {\bf A}\!=\!\frac{({\bf\mu}\!\times\!{\bf\rho})}{\rho^3}\!+
\!\frac{Ze}{\rho}({\bf\omega}\!\times\!{\bf r}_n),
\end{equation} where ${\bf\rho}={\bf r}-{\bf r}_n$.

{\it Nucleus.}---Second component of the H-like ions mentioned
above, {\it i.e.} nucleus having spin $I=1$, is described by the
Proca equation. In the rotating frame $K'$, the wave function
$\phi_\mu$ satisfies the covariant equation
\begin{equation}
\left[g^{\nu\lambda}\nabla_\nu\nabla_\lambda-
\left(\frac{Mc}{\hbar}\right)^2\right]\phi_\mu=0, \label{pr1}
\end{equation}
and the complementary condition
\begin{equation}
g^{\mu\nu}\nabla_\nu\phi_\mu=0, \label{pr2}
\end{equation}
where
$\nabla_\nu\phi_\mu=\partial_\nu\phi_\mu-\Gamma^\sigma_{\nu\mu}\phi_\sigma$
is the covariant derivative of the vector function $\phi_\mu$.
Stationary solutions of these equations imply that
\begin{equation}
\frac{\partial\phi_\mu}{\partial y^0}=-\frac{i}{\hbar c}E\phi_\mu.
\end{equation}
Because $E\sim Mc^2$ in the frame $K'$, the spatial components
$\phi_k$ in (\ref{pr1}) are decoupled from $\phi_0$ and satisfy
the equation
\begin{eqnarray}
-\!\frac{\hbar^2}{2M}\!\sum_i\!\frac{\partial^2\phi_k}{\partial
y^2_i} + i\hbar c\omega_je_{jli}y_l\frac{\partial\phi_k}{\partial
y_i}+ i\hbar c\omega_je_{jki}\phi_i\nonumber\\=(E-Mc^2)\phi_k.
\label{pr3}
\end{eqnarray}
In this approximation Eq. (\ref{pr2}) is reduced to
\begin{equation}
\phi_0= e_{ijk}\omega_iy_j\phi_k\sim\frac{\Omega R_0}{c}\phi_k,
\end{equation}
and hence $\phi_0$ is small compared to the spatial components.
Introducing the vector function
${\bf\Phi}=(\phi_1,\phi_2,\phi_3)$, we can rewrite (\ref{pr3}) in
the matrix form
\begin{equation}
H_n{\bf\Phi}=(E-Mc^2){\bf\Phi},\quad H_n=\frac{{\bf p}^2_n}{2M}-
\hbar{\bf\Omega\cdot}({\bf l}_n+{\bf I}),
 \label{neq}
\end{equation}
where ${\bf p}_n$ is the momentum operator of an ion and ${\bf
l}_n=({\bf r}_n\!\times{\bf p}_n)/\hbar$. The spin operator $\bf
I$ involves the matrices $3\times 3$ in the Decart basis
\cite{Var}.

It should be noted that the spin-rotation couplings in (\ref{eeq})
and (\ref{neq}) are the direct consequence of the covariant
differentiation of spinor and vector functions. If we use in the
flat Dirac and Proca equations solely the spatial coordinate
transformation from $K$ to $K'$, the Hamiltonians $H_e$ and $H_n$
will acquire the Coriolis terms only.

{\it Stored hydrogen-like ion.}---Now we consider the two-particle
Hamiltonian $H=H_e+H_n$ and the corresponding wave equation with
the twelve component wave function. Let us transform this
Hamiltonian to the new variables ${\bf R}=(M{\bf r}_n+m{\bf
r}_e)/(M+m)$ and ${\bf r}={\bf r}_e-{\bf r}_n$. The corresponding
conjugate momenta are ${\bf P}={\bf p}_e+{\bf p}_n$ and ${\bf
p}={\bf p}_e-m{\bf P}/(M+m)$. It is easily proven by direct
calculation that ${\bf l}_e+{\bf l}_n={\bf l}+{\bf L}$, where
${\bf l}$ and ${\bf L}$ are the angular momenta in the variables
${\bf r}$, ${\bf p}$ and ${\bf R}$, ${\bf P}$, respectively.
Complications arise due to the Dirac and Proca differential
equations having different order. As a result, the variables ${\bf
R}$ and ${\bf r}$ do not separate and the former cannot be called
as a center-of-mass coordinate. However, as will be shown below,
the variable ${\bf R}$ can be regarded as the ion center-of-mass
coordinate to an accuracy of $m/M$. Therefore, the Hamiltonian $H$
is expressed as a sum of different components
\begin{equation}
H=H_{cm}({\bf R})+H_0({\bf r})+V_1({\bf r})+V_2({\bf r}, {\bf R}).
\label{ham}
\end{equation}
The first term
\begin{equation}
H_{cm}=\frac{M{\bf P}^2}{2(M+m)^2} -\hbar({\bf\Omega\!\cdot\!\bf
L})-U({\bf R}-{\bf R}_0) \label{cm}
\end{equation}
is the center of mass Hamiltonian with the $\delta$-like potential
$U$ which provides the localization of ion on the stable orbit at
the ring. Relative motion is described by the Hamiltonian
\begin{equation}
H_0=c(\hat{{\bf\alpha}}\cdot{\bf p})\!-\!\frac{Ze^2}{r}\!+
\!\mu_N\frac{eg_I}{r^3}\hat{{\bf\alpha}}\cdot({\bf I}\!
\times\!{\bf r})\!+\!\hat{\beta}mc^2, \label{Hion}
\end{equation}
which includes hyperfine interaction. In the standard
approximation, we label its eigenstates by the quantum numbers
$n,l,j,I$, which will be denoted latter as $\nu$, $F$ and $M_F$,
where ${\bf F}={\bf j}+{\bf I}$ is the total angular momentum. The
third part of $H$,
\begin{eqnarray}
V_1({\bf r})\!=&-&\hbar({\bf\Omega}\cdot{\bf F})-
\frac{Ze^2m}{c(M+m)r}\,\hat{{\bf\alpha}}\cdot({\bf\Omega}\times
{\bf r})\nonumber\\
&-&2\mu_N\frac{eg_I}{c r}({\bf\Omega}\cdot{\bf I}) +\frac{{\bf
p}^2}{2M}, \label{H3}
\end{eqnarray}
includes the perturbations of relative motion resulting from
spin-rotation coupling, the effect of rotation on electron-nucleus
interaction and the "recoil" term. Finally, the potential
\begin{equation}
V_2({\bf r},{\bf R})=-\frac{({\bf P\!\cdot\!\bf p})}{M+m}+
\frac{mc}{M+m} (\hat{{\bf\alpha}}\cdot{\bf P})+ \frac{Ze^2}{cr}
\hat{{\bf\alpha}}\cdot({\bf\Omega}\!\times\!{\bf R}) \label{H2}
\end{equation}
describes the coupling of two degrees of freedom.

The energy spectrum of (\ref{ham}) is found in the adiabatic
approximation. Center of mass motion (slow variable ${\bf R}$) is
described by the Hamiltonian ${\cal H}=H_{cm}({\bf
R})+\varepsilon({\bf R})$, where $\varepsilon$ are the eigenvalues
of the Hamiltonian $H_0({\bf r})+V_1({\bf r})+V_2({\bf r}, {\bf
R})$ involving the swift variable ${\bf r}$ and the parameter
${\bf R}$. The last problem can be solved with perturbation theory
of first order for $V_1$ and second one for $V_2$. The resulting
Sch\"{o}dinger equation describing slow motion is
$$
\left[\frac{M}{2(M+m)^2}\sum_k\left(1+\frac{m}{M}A_k\right)P^2_k
-\hbar({\bf\Omega\!\cdot\!\bf L})-U({\bf R}-{\bf R}_0)\right.
$$
\begin{equation}\left. - \hbar\Omega M_F-
\frac{m}{2}\Omega^2\sum_{k=1,2}B_kX^2_k \right]\Psi({\bf R})
=E\Psi({\bf R}), \label{slov}
\end{equation}
where the dimensionless coefficients $A_k$ and $B_k$ involving the
matrix elements of electron operators are of order 1 for
$Z\alpha\sim 1$, $\alpha$ being the fine-structure constant. They
depend on the quantum numbers $\nu$, $F$ and $M^2_F$. In obtaining
(\ref{slov}) we drop terms of order $(m/M)\hbar\Omega M_F$, which
are negligible compared with the main term of spin-rotation
coupling. The Hamiltonian ${\cal H}$ of eq. (\ref{slov}) allows to
find
\begin{equation}
\dot{{\bf R}}=\frac{i}{\hbar}[{\cal H},{\bf R}]= \frac{{\bf
P}}{M+m}-({\bf\Omega}\times{\bf R})+{\cal
O}\left(\frac{m}{M}\right),
\end{equation}
which is approximately the ion center of mass velocity in the
rotating frame. Considering $\ddot{{\bf R}}$ we get the
approximate equation of motion in the rotating frame with the
Coriolis and the centrifugal forces \cite{Landm}. Thus, the
variable ${\bf R}$ is the center of mass coordinate with high
precision.

The last term in the left hand side of (\ref{slov}) is equal
$(m/2)B_1(\Omega R_0)^2$ due to localization of ion, while the
first term is reasonable to estimate by using the momentum spread
$\delta P$ of ion in the ring. Finally, the energy of the
hyperfine state $\nu,F,M_F$ is approximately
\begin{eqnarray}
E_{\nu F M_F}=E^{(0)}_{\nu F
M_F}-\frac{m}{2M^2}A_\nu(F,M_F)(\delta P)^2\nonumber\\
-\frac{m}{2}B_\nu(F,M_F)(\Omega R_0)^2-\hbar\Omega M_F,
\label{eig}
\end{eqnarray}
where $E^{(0)}$ is unperturbed energy. It is seen that the
coupling of center of mass and relative motions causes the
hyperfine level  to be split in components $|M_F|=F,F-1,...$,
while rotation breaks the degeneracy completely and split magnetic
sub-levels. The levels with $F=j=1/2$ are only shifted and split
in the projections $M_F=\pm 1/2$. Let us note that the first term
in the Hamiltonian (\ref{H3}) can be obtained by using the
transformation (\ref{trans}) with $\bf F$ instead of $\bf j$. This
method has been used earlier in ref. \cite{Silv1}.

The ground state $1^2S_{1/2}$ of the $^{140}{\rm Pr}^{58+}$ and
$^{142}{\rm Pm}^{60+}$ ions having nuclear spin $I=1$ involves the
two hyperfine states $F=1/2$ and $F=3/2$ separated in energy by
approximately 1 eV. According Eq.~(\ref{eig}), the expected
splitting of the lowest $F=1/2$ state is
\begin{equation}
\Delta=\frac{\Omega}{2\pi}\approx\frac{1}{2}(\nu_x+\nu_y)=4.5\
{\rm MHz},  \label{split}
\end{equation}
where $\nu_x$ and $\nu_y$ are the frequencies of the transverse
betatron oscillations, which are determined by the quantities
$Q_x=2.27$, $Q_y=2.23$ and revolution frequency $\nu\approx 2$ MHz
($Q_i=\nu_i/\nu$) typical for the ESR. The shift of the lowest
state is of order $m(\Omega R_0)^2\sim 10^{-6}mc^2$ for $R_0\sim
1$ cm because second term in (\ref{eig}) is small compared to
third one if the momentum spread $\delta P\sim 10^{-7}Mc$.

Electric dipole transitions between the $M_F=\pm 1/2$ sub-states
are forbidden and the calculated mean lifetime of a magnetic
dipole transition is $1.4\cdot 10^{25}$ s. Thus, the natural
line-width of the upper $M_F=-1/2$ sub-state is negligible in
comparison with the splitting (\ref{split}). Even so, the
splitting can not be resolved with the precision laser
spectroscopy developed at the ESR \cite{Kla,See}, because the
Doppler width of the line corresponding to the transition between
the ground state components of the hyperfine structure is
\begin{equation}
\delta\nu=\frac{\nu_0\delta\beta}{1-\beta^2}\sim 100\ {\rm MHz}
\end{equation}
for the relative velocity spread $\delta\beta/\beta\sim 10^{-7}$
and the transition frequency $\nu_0\sim 10^{3}$ THz. Entirely
different type of experiments are needed to observe the
rotationally induced splitting of the order of several megahertz.

To observe the level splitting of hydrogenic Rydberg states,
Silverman \cite{Silv} proposed to use the quantum beat
spectroscopy (see e.g. \cite{Silv2}). The bunched-beam technique
developed in the GSI heavy ion experiments \cite{See} may be
supplemented by the quantum beat spectroscopy. It is convenient to
take the H-like ion having nuclear spin $I=1/2$ and the 1S$_{1/2}$
ground state with $F=0$. However, the observation of the
modulation of the fluorescent light from the transition between
the upper hyperfine state $F=1$ and the ground one is complicated
by the features of ion storing in the ESR. Ions circulating in the
ring are exposed by the fringe transverse (in the plane of the
main orbit) magnetic field of bending magnets. This alternating in
time field causes the repopulation of the magnetic sub-states
because the field strength $H\sim\hbar\Omega/\mu_0=3.2$ G ($\mu_0$
is the Bohr magneton) is enough to mix these states \cite{mix}.
The effects of time varying fields on quantum beats require a
separate theoretical study.

It is interesting to note that the radiative energy loss of
relativistic heavy ions circulating in storage rings is negligible
compared to loses due to collisions with residual gas atoms (see
e.g. \cite{Fran}). Thus the fluorescent light from one bunch will
involve beat signal plus background mostly from ionization and
excitation of residual gas atoms. This background has been
observed in ref. \cite{See}

{\it Ion Interferometry.}---The time-resolved spectroscopy of a
single ion at the ESR along with peculiar properties of the H-like
$^{140}{\rm Pr}$ and $^{142}{\rm Pm}$ ions provide a unique
opportunity for applying the atomic interferometry to resolve the
rotationally induced splitting (\ref{split}). This technique,
which is a prototype of the Stern-Gerlach experiment, successfully
used in a series of the high-precision measurements of the Lamb
shift in a hydrogen atom \cite{Sok}.

According to ref.\ \cite{Ivan}, the EC-decay is forbidden from the
state 1S$_{1/2}$, $F=M_F=1/2$ of these ions due to the neutrino
left-handed helicity. Let us denote this state by $\psi_+$ and the
$F=1/2,M_F=-1/2$ one by $\psi_-$. These states are mixed by a
transverse magnetic field as stated above. A localized magnetic
field is used as a "flipper" mixing the states $\psi_+$ and
$\psi_-$. The interferometer consists of the two flippers I and II
separated by the free field interval of a length $L$. The
evolution of the two-level ion passing the flipper $i$ is
described by the unitary matrix
\begin{equation}
{\cal U}_i=\left(\begin{array}{cc} D_i& -A^*_i\\
A_i& D_i \end{array}\right),\ D_i=\sqrt{1-|A_i|^2},
\end{equation}
where $A_i$ is the corresponding transition amplitude. Between the
flippers, the adiabatic evolution is given by the diagonal matrix
\begin{equation}
{\cal S}(\tau)\!=\!\left(\begin{array}{cc} 1& 0\\
0 &\exp{[(i\Omega -\lambda/2)\tau]} \end{array}\right),\
\tau\!=\!L\sqrt{1-\beta^2}/V,
\end{equation}
where $\lambda$ is the rate of the EC-decay. The ion velocity in
the laboratory frame is approximately equal $V$ because its
velocity relative to the inertial frame $K$ is of the order
$\Omega R\sim 10^7$ {\rm cm/s. The value $V$ is used also to take
into account the Lorentz contraction of the interval $L$.

Let an ion enter the interferometer at the time $t$ in the mixed
state $a(t)\psi_+ +b(t)\psi_-$. After passing sequentially the
flippers I and II, its state is changed into $a(t+\tau)\psi_+
+b(t+\tau)\psi_-$, where the new mixing coefficients are connected
with old ones by
\begin{equation}
\left[\begin{array}{c}a(t+\tau)\\b(t+\tau)\end{array}\right]=
{\cal U}_2{\cal S}(\tau){\cal U}_1
\left[\begin{array}{c}a(t)\\b(t)\end{array}\right]. \label{iter}
\end{equation}
The probability of the ion decay in some region beyond the flipper
II is proportional to $|b(t+\tau)|^2$. This value involves
interference terms because the two states evolve between the
flippers with the phase difference $\Omega\tau$. However, in the
technique of the single ion interferometry,  an interference
pattern  can not be observed if the amplitudes $a(t)$ and $b(t)$
are random owing to the chaotic repopulation of these states in
the ring.

To circumvent this obstacle an ion before entering the
interferometer must be polarized in the state $\psi_+$ in a rf
resonance cavity by means of the Lamb and Retherford method
\cite{Lamb}. With the polarized ion eq. (\ref{iter}) gives
\begin{eqnarray}
|b(t+T)|^2=D^2_1|A_2|^2+|A_1|^2D^2_2e^{-\lambda\tau}+\nonumber\\
2|A_1|D_1|A_2|D_2e^{-\lambda\tau/2}\cos(\Omega
\tau+\varphi_1-\varphi_2),
\end{eqnarray}
where $\varphi_1$ and $\varphi_2$ are the phases of the amplitudes
$A_1$ and $A_2$. As follows from this equation, the period of
spatial oscillations is $L_0=V/(\Delta\sqrt{1-\beta^2})\approx 68$
m, well above the field-free straight section of the storage ring.
The difficulty can be overcome by using the method of ref.
\cite{Sok}, which involves the following measurements: \\
(i) For the fixed distance $L$ (time $\tau$) and the flipper
magnetic fields ${\bf H}_1$ and ${\bf H}_2$, the number $N$ of
EC-decays per second is collected in some spatial interval beyond
the flipper II (which may be wide enough for the considered ions
because the decay length $L_d=V/(\lambda\sqrt{1-\beta^2})$ is well
above the circumference of the ring). \\
(ii) After the field in the flipper II is reversed (${\bf
H}_2\rightarrow -{\bf H}_2$), the value $N'$ is obtained. \\
(iii) The two measurements allow to find the difference $\Delta
N(L)=N-N'$. \\
(iv) Analogously, the differences $\Delta N(L+\Delta L)$ and
$\Delta N(L-\Delta L)$ corresponding to the altered intervals are
acquired. \\
From these measurements, one gets
\begin{equation}
\Omega\!=\!\frac{V}{\Delta
L\sqrt{1\!-\!\beta^2}}\arccos{\frac{\Delta N(L\!+\!\Delta
L)+\Delta N(L\!-\!\Delta L)}{2\Delta N(L)}},
\end{equation}
if $\lambda\tau\ll 1$. The main uncertainty in determing $\Omega$
comes from the measurements of $\Delta N$ because the velocity
spread of ions, $\delta V/V\sim 10^{-7}$, is very small.

{\it Conclusion.}---The coupling of the particle spin with
rotation reveals its rotational inertia. Therefore the coupling
has to be considered by unified way in the framework of general
relativity. We derive this inertial effect for the
spin-$\frac{1}{2}$ and the spin-1 particles by using the general
covariance principle. The result of calculations shows that spin
incorporates into gravity as an ordinary angular momentum.
Correspondingly, the gyrogravitational ratio of {\it bare}
elementary particles is equal unity.

Observational evidence for spin-rotation coupling includes the
experiments of two types. The interference experiments provide the
direct measurement of the gyrogravitational ratio. A point of
interest in such experiments, does this ratio for elementary
particles differ from unity? The neutron interferometry
experiments were proposed by Mashhoon \cite{Mash,Mash1}, but has
not yet been performed. To observe spin-rotation coupling the
storage ring based experiment with H-like ions may be used. These
experiments are less precise than the interference ones because of
the fine systematic effects of the ion internal structure.

Due to the betatron oscillations of a stable ion orbit, an ion has
ability to rotate around the direction of its motion in the ring
with the approximately doubled circulating frequency. In the
rotating reference frame internal dynamics of the ion is described
by using the generally covariant equations. It has been found that
the degenerate magnetic sub-states of a hyperfine structure are
split due the interaction of rotation with the ion total angular
momentum and the effect of the curved space-time on the
electromagnetic interaction electron and nucleus. The splitting of
the ground hyperfine state $1^2S_{1/2},F=1/2$ of the $^{140}{\rm
Pr}^{58+}$ and $^{142}{\rm Pm}^{60+}$ ions circulating in the ESR
is about 4.5 MHz and is caused mainly by spin-rotation coupling.
It can be detected with the method of ion interferometry. This is
conceptually new method in which the coherent superposition of two
electron states of the ion is destroyed by the orbital electron
capture decay of nucleus. One may hope that with the ion
interferometry installed at high-Z ion storage rings new high
precision experiments will be available.

\acknowledgments The author is grateful to  A. Dolinski and V. L.
Ushkov for helpful information concerning the stable ion orbits in
storage rings. I am grateful also to S. T. Belyaev, A. A.
Korsheninnikov and Yu. N. Novikov for a critical reading of the
manuscript. The work was supported by the Grant NS-3004.2008.2.


\begin{thebibliography}{99}
\bibitem{mix} A repopulation of levels has not been observed  in
the GSI heavy ion experiments \cite{Kla,See} because the magnetic
field $H\sim h\nu_0/\mu_0\sim 10^4$\ T is necessary for the two
ground state hyperfine levels of heavy H-like ions to be fully
mixed.
\end{thebibliography}
\end{document}